# Planning & Optimization of Wireless LAN's through Field Measurements

Puneet Kumar Mongia[1], Dr. B. J. Singh[2].

1. School of VLSI Design & Embedded Systems, National Institute of Technology, Kurukshetra, Haryana, India. e-mail: mongiap@gmail.com,
2. Department of Electronics & Communication Engineering, National Institute of Technology, Kurukshetra, Haryana, India. e-mail: brahmjit@nitkkr.ac.in.

**ABSTRACT**

In this paper, the field measurements of signal strength taken at the frequency of 2432 MHz in indoor & outdoor environments are presented and analyzed. The received signal levels from the base station were monitored manually. Total coverage area considered for the measurement campaign consisted of a mixture of different propagation environments. Based on the experimental data obtained, path loss exponent and standard deviation of signal strength variability are derived. It is shown that the values of these parameters vary from region to region in the coverage area. The analysis is purely statistical & is compared with the existing propagation model to determine the path loss exponent & deviation.

## 1. INTRODUCTION

Wireless technologies are very popular for the flow of information. IEEE 802.11 technologies have started to spread rapidly, enabling consumers to set up their own wireless networks [1]. Also with the increasing use of mobile computing devices such as PDAs, laptops, and an expansion of Wireless Local Area Networks (WLAN), there is growing interest in optimizing the WLAN infrastructure so as to increase productivity and efficiency in various colleges and office campuses with carrying out a cost effective infrastructure model [8]. So it has become essential to understand the propagation characteristics for a proposed WLAN before deployment. One of the most important reasons of this is security. Predicting how far a signal can go before installation will ensure that a connection cannot be made in areas where it is not desired. The strength, range and coverage area of an access point is strongly affected by its positioning in reference to its environment. To avoid undesired connection in places where it should not be made, it is necessary to carefully predict the signal strength, range, coverage and the correct placement of base stations [2].Various mechanisms like reflection, diffraction, refraction, scattering & absorption also affects the strength of the signal [3].The strength of our existing WLAN for indoor & outdoor access points can be analyzed by statistical field measurements. This paper demonstrates the analysis of the existing wireless network in indoor & outdoor environments by examining the field measurements of received signal strength (RSS). The data taken for the analysis is used to calculate the path loss exponent & standard deviation of the location [3]. The frequency of the radio signal analyzed is 2432 MHz. The received signal levels were monitored from the access point (AP) or base station (BS) manually. Total coverage area chosen for the measurement campaign consisted of a mixture of significantly different propagation environments [4].

## 2. RSS MEASUREMENTS OR SITE SURVEY

### 2.1 Area Chosen

We performed experiments using a subset of the existing indoor wireless infrastructure in the Electronics & Communication Engineering Department at National Institute of Technology, Kurukshetra for indoor analysis using the indoor access point and outside UIET Kurukshetra University Kurukshetra, for outdoor analysis using outdoor base stations. We considered our indoor measurements at 4 locations, two rooms at ground floor in which one room containing the access point and other room is just next to it. The other two locations were the corridors at the ground and the 1st floor. For outdoor measurements we have considered 2 locations one outside the UIET building covering the front boundary and 2nd is in the park outside the Geophysics Department of Kurukshetra University which is approx. 12-15 meter distance from the access point. A total of 25-30 readings were taken at each location and the area covered varies from 8 to 12 meters. Figure 1 & 2 shows the locations on google map.

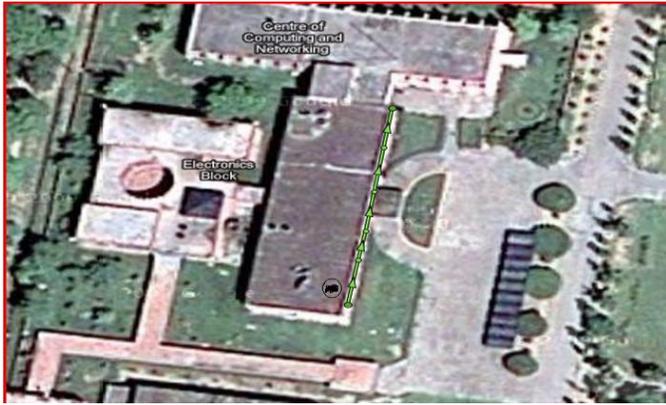

**Figure 1: G-Map view of readings taken at NITK inside the ECE BLOCK building. The circle shows the approx. location of the access point inside the building in ground floor.**

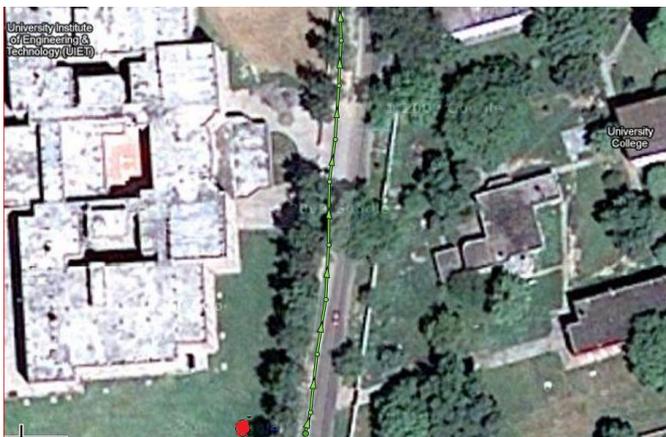

**Figure 2: G-MAP view of the location at UIET where readings were taken. The small red circle determines the location of base station or access point.**

### 2.2 Data Collection Tools

For site surveying, generally wireless sniffing tools are used to sniff wireless packets from an ad-hoc network setup using an access point. There are varieties of open source wireless sniffing tools available on the internet [8]. In our analysis, Passmark Wirelessmon Software was used to provide the values of received signal strength. It also gives the MAC (Media Access Layer) address of the access point along with the frequency value and standard of the AP. Its trial version is available on the internet. It also plots the graph between signal strength and time. The software has the sensitivity of 200 dBm (milli decibel) [5].The figure 3 shows the snapshot of the software.

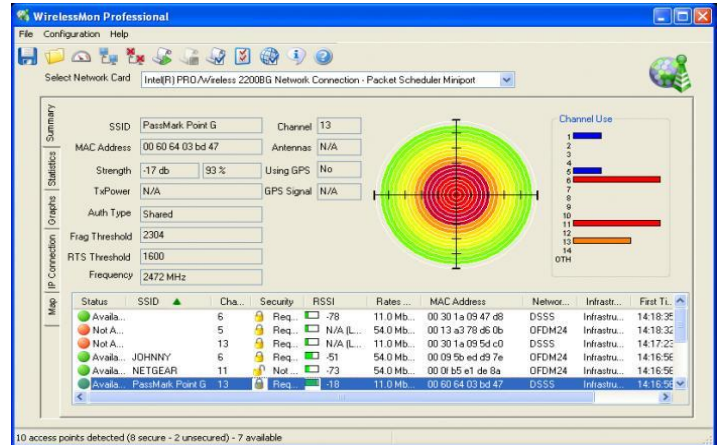

**Figure 3: Snapshot of Passmark Wirelessmon Software**

Another software called as Ekahau Heatmapper was used to determine the coverage region of an access point graphically by considering a reference location of the AP on the software. If the map of the location is available then the map can be loaded in the software and with the help of that the site survey of the location can be studied and also the signal strength can be determined [9].MATLAB 7.0.4 software was used for the data analysis of the readings in terms of plotting of graphs and curve fitting [6] [10].

Our client or receiver device was a laptop of Dell Company of model Vostro 1510 having a wireless NIC (Network Interface Card) of INTEL(R) PRO/WIRELESS 3945ABG NETWORK CONNECTION. It supports WLAN protocols of type IEEE 802.11 a/b/g. It was having the maximum sensitivity of 95 dBm [12] [14].

## 3. INTRODUCTION TO PROPAGATION MODELS

### 3.1 Importance of propagation models

The wireless channel places fundamental limitations on the communication systems because of its dynamic nature. The transmission path between the transmitter and the receiver can vary from simple line of sight to one that is obstructed by trees, walls, floors etc. This can cause the signal strength of two points that are equidistant from the access point to be entirely different. Propagation models focus on predicting the average signal strength that may be received at a particular distance from a transmitter. Thus, it is important to determine the propagation model for the indoor wireless network by taking shadowing into account [3] [8]. In the following sub sections we will discuss about some of these models.

### 3.2 Free Space Path Loss Model

Free space propagation model is used to predict the signal strength at a distance from the receiver when there is no obstruction between the transmitter and the receiver. It is the foundation for all other models. It is derived from Friis's free space equation given by:-

$Pr(d) = (Pt*Gt*Gr*W^2) / (4pi)^2 * d^2 * L$, Where,

$Pr(d)$ = power received at a distance 'd' from transmitter.

$Pt$ = transmitted power.

$Gt, Gr$ = transmitter and receiver antenna gains respectively.

L = largest antennae dimension.
W=Wavelengh of the signal [3].
The path-loss represents the attenuation the signal undergoes as it propagates through the medium. It is given by,
PL (dB) = 10log (Pt/ Pr)
= -10log (Gt*Gr*W^2/ (4pi) ^2*d^2*L [3] [11].

### 3.3 Log Distance Path Loss Model

The free space propagation model is a theoretical model; not applicable to real life situations. Log distance path loss model is a practical path loss model which is based on the fact that the received power decreases logarithmically with distance. The average path loss for an arbitrary Tx-Rx separation is given by the following expressions,
PL (dB) is proportional to (d/d0) ^ n, or
PL (dB) = PL (d0) + 10nlog (d/d0) + X (sigma)
Where, d=distance of the receiver from the access point or the reference point in foot or meters, here n = path loss exponent, which can be used as a measure of path loss in a particular area. So if graph is plotted between the log distance & path loss in deceibel than the slope of the linear curve will give the value of n.
Here, PL (d0) = power received at reference distance 'd0' & X (sigma) represents a normal random variable in dB having a standard deviation of sigma dB.
In terms of received power, the same propagation model can be re-written as,
Pr (dB) = Pr (d0) - 10nlog (d/d0)   [3].
This model can also be used for outdoor propagation for small distances between transmitter and receiver [8]. On the basis of above equations, graphs were plotted between the power loss & distance on log scale. Thereafter, the table is drawn showing the path loss & deviation of each location.

### 3.4 Indoor Environment Analysis

For the indoor analysis the access point or base station used was having the following specifications:
SSID – ECE_MICRO
RF Transmitting Power - 200 mW (23 dBm),
Vendor – Semindia Systems Private Limited,
Model - WA3002G4,
Support - IEEE 802.11g (OFDM 24)    [13].
The passmark wirelessmon software gave the received power reading at the current location of the receiver which determined the path loss on that particular point using the log distance model [5] [3].
The following subsections illustrate the set of locations along with the figures showing the variation of path loss with respect to the distance from the access point or reference point. On the basis of the readings obtained a linear fit of the graph was drawn to determine the approximate path loss & deviation using the MATLAB tool.

*3.4.1  Room 1*

This was the seminar room on the ground floor where the AP was located. The readings were taken directly from the AP and approaches away from the AP in straight direction. The total numbers of samples taken were 22.
As we have discussed that the power of the transmitter (i.e. AP) is 200 mW, so in dBm it will be=10*log (200m/m) =23 dBm(approx.). Now we know that Path Loss in dB is given by the difference between the power of transmitter in dBm and power of receiver in dBm.
So PL(dB)=Pt(dBm)-Pr(dBm)   [3].
The received power is also known as RSSI (Received Signal Strength Indication). Figure 4 shows the variation of the path loss (in dB) of this location with respect to the distance from the AP varying in feet on log scale. The straight line shows the linear fit of the readings.

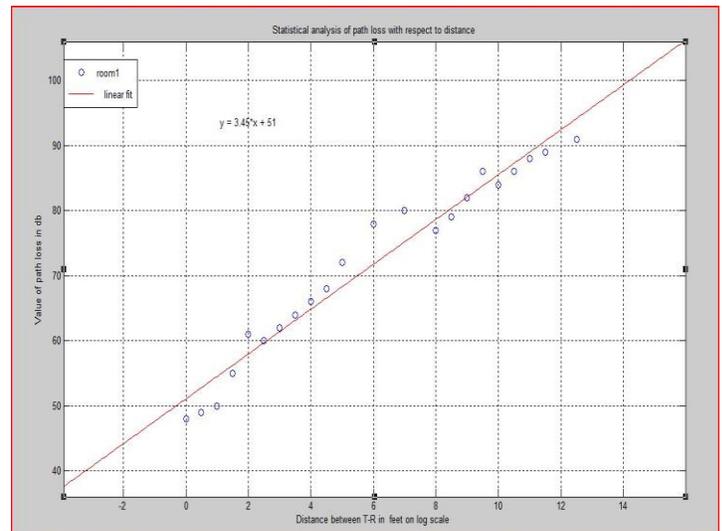

**Figure 4: Showing the variation of path loss in dB along y axis with respect to the varying distance from the AP in feet on log scale along x axis. Straight line shows the linear fit of the readings.**

*3.4.2  ROOM 2*

This was the room just next to the room where the access point was located. The readings were taken from the door and approaches away from the door into the room. The distance of the AP from the door of the room was approx. 9 feet. The total samples taken were 19.

Figure 5 shows the variation of the path loss (in dB) of this location with respect to the distance from the reference point (door) varying in feet on log scale. The curved line shows the linear fit of the readings.

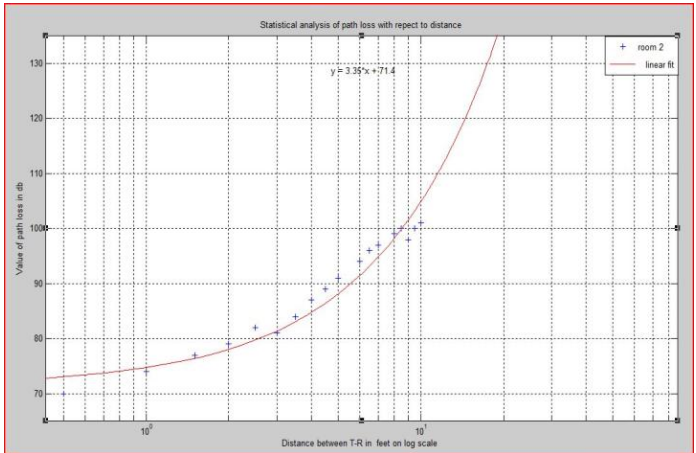
**Figure 5:** Showing the variation of path loss in dB along y axis with respect to the varying distance from the reference point in feet on log scale along x axis. The curved line shows the linear fit of the readings.

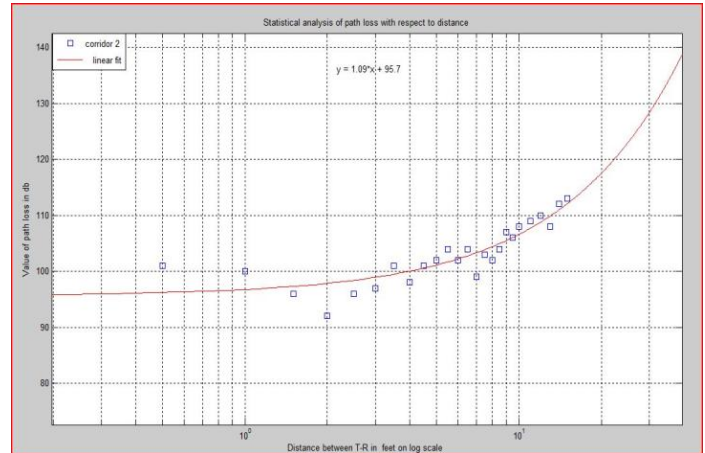
**Figure 7:** Showing the variation of path loss in dB along y axis with respect to the varying distance from the reference point in feet on log scale along x axis. The curved line shows the linear fit of the readings.

### *3.4.3 CORRIDOR 1 (GROUND FLOOR)*

This was the corridor at the ground floor. The AP was located almost at the extreme left of the reference point. The readings were taken in the whole corridor. The perpendicular distance of the AP from the starting point was approx. 16 feet. The total samples taken were 28.

Figure 6 shows the variation of the path loss (in dB) of this location with respect to the distance from the reference point varying in feet on log scale. The curved line shows the linear fit of the readings.

### *3.4.5 OVERALL INDOOR LOCATIONS*

Here height of transmitting AP was 8 feet for all locations and height of receiving point taken was 3.5 feet for room 1, 2 & corridor1 and 20 feet for corridor 2 with respect to ground level.

Figure 8 shows the variation of the path loss (in dB) of all the locations with respect to the distance from the reference point varying in feet on log scale.

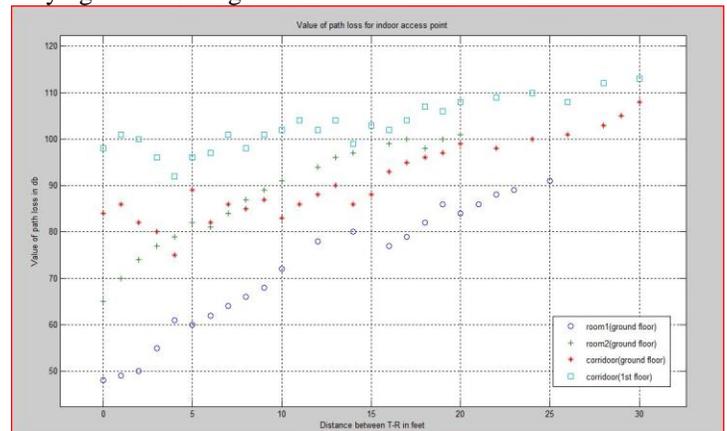
**Figure 8:** Showing the variation of path loss in dB along y axis with respect to the varying distance from the reference point in feet on log scale along x axis for all the locations.

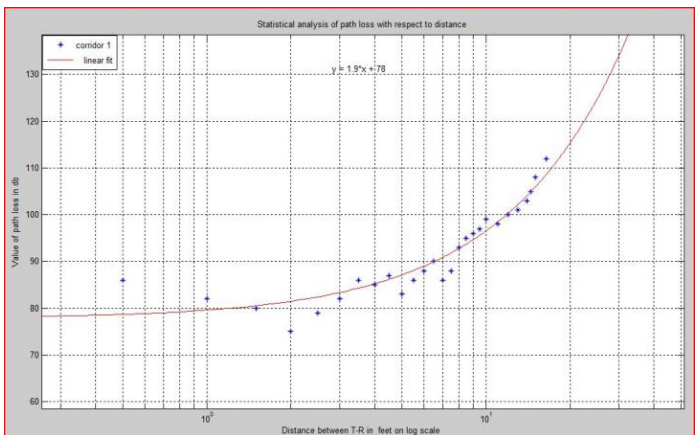
**Figure 6:** Showing the variation of path loss in dB along y axis with respect to the varying distance from the reference point in feet on log scale along x axis. The curved line shows the linear fit of the readings.

### *3.4.4 CORRIDOR 2 (FIRST FLOOR)*

This was the corridor at the 1st floor. The AP was located just below the corridor. The readings were taken from the starting of corridor. The total samples taken were 28.

Figure 7 shows the variation of the path loss (in dB) of this location with respect to the distance from the reference point varying in feet on log scale. The curved line shows the linear fit of the readings.

### 3.5 Outdoor Environment Analysis

For the indoor analysis the access point used was having the following specifications:
SSID – UIET-WN2
RF Transmitting Power - 200 mW (23 dBm),
Vendor – Wavion Wireless Limited,
Channels – 1, 5, 6,7,11
Frequency – 2432 MHz.
Support - IEEE 802.11g (OFDM 24)    [14] [15].
The RSSI values were obtained using the passmark wirelessmon software as we did for indoor analysis [5] [3].
The following subsection illustrates the set of locations along with the figures showing the variation of path loss with respect to the distance from the access point or reference point.

### 3.5.1 Location 1

This was the nearest location along the AP. The readings were taken just outside the UIET building. The readings were taken directly from the access point and approaches away from the AP in straight direction. The total samples taken were 26.

Figure 9 shows the variation of the path loss (in dB) of this location with respect to the distance from the AP varying in feet on log scale. The curved line shows the linear fit of the readings.

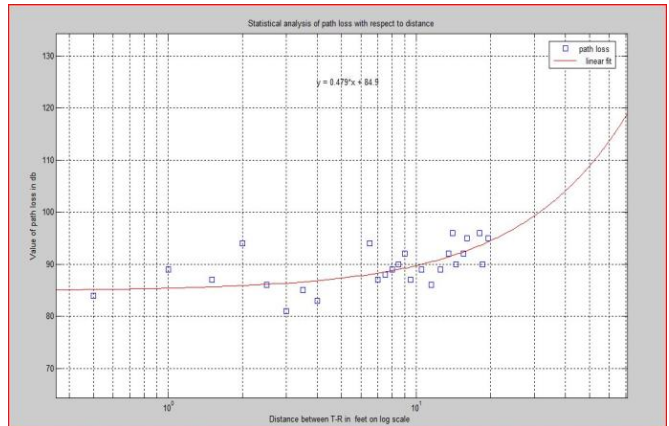

**Figure 9:** Showing the variation of path loss in dB along y axis with respect to the varying distance from the AP in feet on log scale along x axis. The curved line shows the linear fit of the readings.

### 3.5.2 Location 2

This was the location situated at a distance of 10-12 meter from the AP. The readings were taken in front of the Geophysics department building. The distances were taken such that we were approaching away from the AP. The total samples taken were 20.

Figure 10 shows the variation of the path loss (in dB) of this location with respect to the distance from the reference point varying in feet on log scale. The curved line shows the linear fit of the readings.

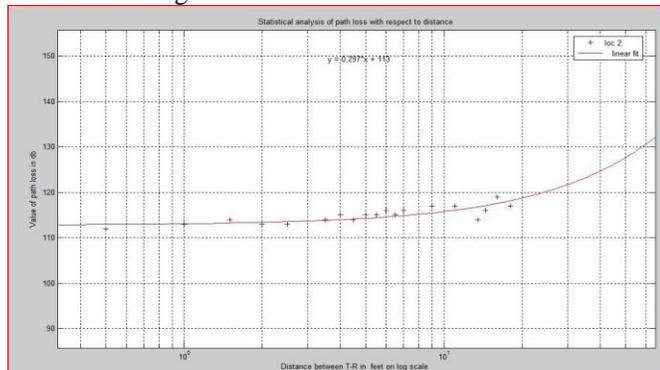

**Figure 10:** Showing the variation of path loss in dB along y axis with respect to the varying distance from the reference point in feet on log scale along x axis. The curved line shows the linear fit of the readings.

## 4. RESULTS

From the sampled readings & linear curve fit for all locations, we have obtained the path loss & standard deviation of each location which are listed below in Table 1.

| Location | Path Loss Exponent | Deviation (dB) |
|---|---|---|
| Room1 | 3.45 | 13.92 |
| Room2 | 3.36 | 11.10 |
| Corridor 1 | 1.88 | 9.45 |
| Corridor 2 | 1.09 | 5.25 |
| Location 1 | 0.48 | 4.32 |
| Location 2 | 0.30 | 4.32 |

**TABLE 1:** Path Loss & Standard Deviation for all locations.

With the help of Ekahau Heatmapper software we can calculate the coverage area of a particular access point. Figure 11 shows the coverage region of the outdoor access point by considering the access point at the origin [9].

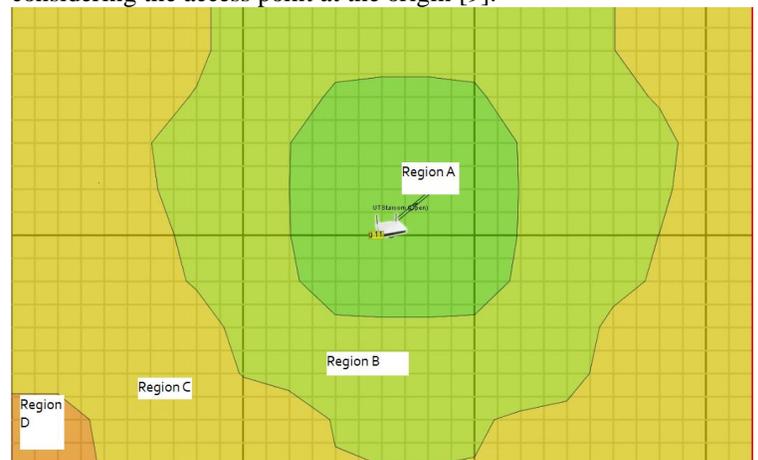

**Figure 11:** Showing the coverage region for the outdoor access point.

Based on the coverage region figure we have divided the coverage area into four regions with each region having its effective signal strength. Table 2 accounts for the range & signal strength of each region.

| Region | Effective Range | RSSI (dBm) |
|---|---|---|
| A | 2-4 METER | -56 TO -48 |
| B | 4-10 METER | -64 TO -56 |
| C | 10-25 METER | -72 TO -64 |
| D | MORE THAN 25 | -80 TO -72 |

**TABLE 2:** Coverage region of an AP.

From these readings we can say that the strength of the signal varies as we go away from access point. Also the signal strength is Strong in region A & B. It is fair in region C & Very Low in region C. So to get the high speed in using internet the user must choose region A or B. To download high memory applications, these two regions are optimum. For using moderate applications region C is also effective. Region D has very low strength and speed will be very low.

## 5. CONCLUSIONS

In In this paper, the field measurement results and statistical analysis of RSS conducted at frequency of 2432 MHz for indoor environment of ECE dept. at NIT Kurukshetra & for the outdoor environment of UIET Kurukshetra University have been reported. The study shows the variability of the propagation characteristic parameters from one environment to another over the coverage area. The path loss exponents for different areas are obtained by utilizing the curve fitting method. The path loss exponents obtained lie in the range from 1.09 to 3.45 for indoor environment & 0.30 to 0.48 for outdoor environment. The standard deviation is a measure of the impreciseness of the area. If, for generic system studies, path loss is taken of simple form depending only on distance but not on details of the path profile, the standard deviation will necessarily be large. So the standard deviation tells us the effective fluctuation rate of the signal which depends not on the distance but on the environment. As we can see from our analysis, deviation is more for indoor environment. The results in the paper may be utilized as reference in the system level simulation for network planning and optimizing the design parameters during the network setup [4]. Before the setup of a WLAN a proper planning about the area, infrastructure, number of access points, number of users, type of modulation, type of application use is required. In the existing setup this type of site survey and statistical analysis can help to improve the quality of the signal strength, by changing the position of the access points & by establishing new access point where the strength of the signal is low. Also by changing the height or the direction of antenna in case of directional antennas can improve the existing wireless LAN setup [7]. The path loss increase not only depends on distance factor but also due to the attenuation produced by the various elements and things present on the location. The path loss increases when the receiver approaches through walls or metal covering due to penetration of the signal. The strength is not so much affected due to the presence of trees and plants on the way. The strength depends on the time of the day; due to the scattering effect [3].Strength also depends on the climate factors. Generally the laptops or the mobiles with WLAN connectivity have the maximum sensitivity varying from -90dBm to -120 dBm, so whenever the RSSI value at any location is more than 10 dBm from the sensitivity value then there is a need of establishing a new access point at that location. To increase the coverage area one way is to increase transmitter power but that is not a good choice as it will increase the path loss as well. So it is better to change the position of the AP or use better modulation techniques. Generally the modulation used in this LAN is OFDM (Orthogonal Frequency Division Multiplexing) but there is a need to use the ADM (Adaptive Delta Modulation) because as we approach away from the access point the strength decreases and signal is very weak, so to make more effective use of the Wireless LAN for far range the modulation used should be adaptive.

## ACKNOWLEDGMENT

Many people helped us through our work and we want to take this opportunity to acknowledge their contribution. We would like to thank all the faculty members of our department & UIET Kurukshetra University for permitting us to do survey in department and also for providing us with necessary information as and when required.

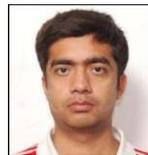
**Puneet Kumar Mongia**, is a final year student of M. Tech. in School of VLSI Design & Embedded Systems at NIT Kurukshetra. His area of interests are Wireless Communication systems, FPGA & VLSI design.

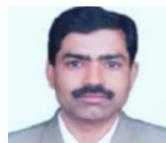
**Dr. B. J. Singh** is a professor at NIT Kurukshetra in Electronics & Communication Engineering Department. He has more than 24 years of experience in the field of Wireless Networks and Mobile Communication, Wireless Sensor Networks & Cognitive Radio. Dr. B. J. Singh has received best Research Paper Award from IE (INDIA).